\documentclass[aps,prl,letter,reprint,amsmath,amssymb,groupedaddress,floatfix,showpacs,showkeys,subeqn]{revtex4-1} 
\usepackage{hyperref}
\usepackage{graphics} 
\usepackage{graphicx} 
\usepackage{subfigure}
\usepackage{url}

\begin{document} 
\author{L. A. Toikka}
\author{K-A. Suominen}
\affiliation{Department of Physics and Astronomy, University of Turku, 20014 Turku, Finland}
\title{The Snake Instability of Ring Dark Solitons in Toroidally Trapped Bose-Einstein Condensates}
\date{\today}
\begin{abstract}
We show that the onset of the snake instability of ring dark solitons requires a broken symmetry. We also elucidate explicitly the connection between imaginary Bogoliubov modes and the snake instability, predicting the number of vortex-anti-vortex pairs produced. In addition, we propose a simple model to give a physical motivation as to why the snake instability takes place. Finally, we show that tight confinement in a toroidal potential actually enhances soliton decay due to inhibition of soliton motion.
\end{abstract}

\maketitle

%\section{Introduction}
The nonlinearity of the equations of motion governing the behaviour of optical systems and ultracold atomic gases allows for interesting solutions such as bright and dark solitons~\cite{Kivshar1998,P&S,Emergent,Greekreview}, which remarkably propagate without dispersion~\cite{Drazin1989} and scatter elastically. In nonlinear optics, this means that soliton light pulses sent through optical fibers propagate without changing shape, whereas in Bose-Einstein condensates and superfluid Fermi gases~\cite{PhysRevA.83.041604}, matter wave solitons correspond to shape-maintaining dips or humps in the atomic density.

In both settings dark solitons have been observed experimentally~\cite{PhysRevLett.69.2503,PhysRevLett.83.5198}, but strictly speaking the stability holds only in one-dimensional systems. In higher dimensions, in general, they collapse into vortex-anti-vortex pairs in 2D (or vortex rings in 3D) through the snake instability~\cite{ZacharyDutton07272001, PhysRevLett.86.2926}, even though dark solitons retain many of their solitonic properties in nearly-integrable systems, for example, in the presence of an external trap~\cite{PhysRevA.68.043613}. The stability and dynamics of dark solitons has been discussed theoretically~\cite{PhysRevA.60.R2665, PhysRevE.49.1657, PhysRevA.72.023609,Kivshar_dynamicsof}, in particular, complex frequencies in the Bogoliubov-de-Gennes (BdG) excitation spectrum have been demonstrated to drive the instability~\cite{PhysRevA.62.053606}.

Cylindrically symmetric systems, bringing forward the concepts of ring bright~\cite{PhysRevLett.97.010403} and dark~\cite{PhysRevE.50.R40, PhysRevLett.90.120403, PhysRevE.66.066611} solitons (RDS), offer an example of a two-dimensional system, where the dynamics can be reduced to a one-dimensional equation, and thereby making the question of stability relevant. It has been observed that filling the notch of a RDS stabilises against the snake instability~\cite{0953-4075-44-19-191003}, a result we explain in this Letter. Specifically, ring traps~\cite{Griffin2008,Baker2009,Sherlock2011,PhysRevA.64.063602} might stabilise the ring dark soliton, but we show that their effect is in fact the opposite.

In this Letter we consider the stability of the RDS in cylindrically symmetric confinement, showing that it is stable as long as the symmetry in the $\theta$-direction is maintained. We show that the type of the simulation grid plays an important role related to the induced breaking of this symmetry. We propose a physical model explaining why the snake instability takes place, and further elucidate the connection between the complex BdG spectrum and the snake instability. Based on this model, we show how the number of vortex-anti-vortex pairs is directly related to the BdG spectrum.

To investigate the properties of the two-dimensional condensate, we assume it is described by a macroscopic wavefunction, $\psi$, which solves the Gross-Pitaevskii equation in the $(x,y)$ plane:
\begin{equation}
\label{GPE2Da}
i \psi_T =  - \nabla^2 \psi + V_{\mathrm{trap}} \psi + C_{2D} |\psi|^2 \psi,
\end{equation}
where $C_{2D} = 4 \sqrt{\pi} Na/(a_{\text{osc}}^{(z)})$, where $N$, $a$, and $a_{\text{osc}}^{(z)}$ are the number of atoms in the cloud, the $s$-wave scattering length of the atoms, and the characteristic trap length in the $z$-direction respectively. We have obtained dimensionless quantities by measuring time, length and energy in terms of $\omega_x^{-1}$, $a_{\text{osc}} = \sqrt{\hbar/(2m\omega_x})$ and $\hbar \omega_x$ respectively, where $\omega_x$ is the angular frequency of the trap in the $x$-direction. This is equivalent to setting $\omega_x = \hbar = 2m = 1$. Similarly in the polar coordinates $(r,\theta)$:
\begin{equation}
\label{GPE2Dapolar}
i \phi_T =  -\phi_{rr} - \frac{1}{r^2} \phi_{\theta \theta} + \left(V_{\mathrm{trap}} - \frac{1}{4r^2}\right)\phi + \frac{C_{2D}}{r} |\phi|^2 \phi,
\end{equation}
where we have used the standard scaling $\phi = \sqrt{r}\psi$. The normalisation of $\psi$ and $\phi$ is to unity. Exact soliton-like solutions for Eq.~\eqref{GPE2Dapolar} are difficult to find, although some can exist for very specific forms of $V_{\mathrm{trap}}$ ~\cite{ths2012}.

The potential term $V_{\mathrm{trap}}$ is the external trap, and we consider cylindrically symmetric traps, for which $\omega_y = \omega_x \equiv \omega_r$. In particular, we consider harmonic ring traps with toroidal potentials
\begin{equation}
\label{SonT:1}
V_{\mathrm{trap}} = \frac{k}{4}(R-R_S)^2,
\end{equation}
where $R_S$ is the radius of the trap minimum, and $k$ is a real nonzero constant. 

To find the ground and excited states and to consider time evolution, we propagate Eqs.~\eqref{GPE2Da} and~\eqref{GPE2Dapolar} in imaginary and real time respectively, using a split-operator Fourier method~\cite{0034-4885-58-4-001}. We consider Cartesian $(x, y)$ and polar $(r, \theta)$ coordinate meshes and for the polar mesh we extend the grid to the negative $r$-axis and demand the wavefunction be antisymmetric about $r = 0$. The ring solitons are generated by printing an appropriate phase step in the radial direction.

We also explore the Bogoliubov quasiparticle (or normal eigenmode) excitation spectrum, defining the Bogoliubov amplitudes $u$ and $v$ by $\lbrace u, v \rbrace (\textbf{r}) = e^{iq\theta}\lbrace u_q, v_q \rbrace (r) $~\cite{PhysRevA.58.3168} representing a partial wave of angular momentum $q$ relative to the condensate. The Bogoliubov-de-Gennes equations for low-lying modes $\Omega_q$ in dimensionless harmonic oscillator units ($\omega_x = \hbar = 2m = 1$) are
\begin{equation}
\label{BdG_matrix}
\begin{pmatrix} \mathcal{H} & C_{2D} \phi_0^2 \\ C_{2D} \bar{\phi}_0^2 & \mathcal{H} \end{pmatrix} 
\begin{pmatrix} u_q \\ v_q \end{pmatrix} = \Omega_q \begin{pmatrix} u_q \\ -v_q \end{pmatrix},
\end{equation}
where $\mathcal{H} = -\nabla_{r,\theta}^2  + V_{\mathrm{trap}}  + 2C_{2D}|\phi_0|^2 - \mu$, and $\phi_0$ is the cylindrically symmetric self-consistent wavefunction, containing a ring dark soliton. Here $\mu$ is the chemical potential, and $u$ and $v$ satisfy the general condition $(\Omega_q - \bar{\Omega}_{q'})\int d\textbf{r} (\bar{u}_q u_{q'} - \bar{v}_{q'}v_q) = 0$. The instability time is given by $1 / \mathrm{Im}(\Omega_q)$, and if all eigenvalues are real, the condensate is dynamically stable. We employ Lagrange functions~\cite{0305-4470-19-11-013} together with Hessenberg QR iteration~\cite{laug}.

We have observed the snake instability of the ring dark soliton only when the symmetry in the $\theta$-direction is broken. This can happen, for example, by using a rectangular coordinate system $(x,y)$ and a symmetrical potential, or artificially adding an anisotropy in the trapping potential and using a (symmetrical) polar coordinate system $(r,\theta)$. In particular, in our simulations the ring dark soliton is a stable object as long as this $\theta$-symmetry is maintained. The previous results with clear decay into a vortex-anti-vortex necklace in the literature~\cite{PhysRevLett.90.120403} follow because a rectangular grid was used. It is clear that in such a situation the vortex-anti-vortex pairs also appear in multiples of four because the simulation setup has reflection symmetry about two orthogonal axes (what happens in one quadrant is copied 4-fold).

When there exists symmetry in the $\theta$-direction, the Gross-Pitaevskii equation governing the behaviour of the system becomes effectively one-dimensional in the radial direction. In the rectangular case, for the equation to become effectively one-dimensional, one would need an infinite system that is homogeneous in one direction (or in 3D homogeneous in $x$ and $y$ if the dark soliton is in the $z$-direction). In Ref.~\cite{PhysRevA.60.R2665} it was shown that such a planar dark soliton is stable as long as the transverse dimensions are restricted enough to make the system quasi-one-dimensional, but a similar quasi-one-dimensionality is achieved by the transversal symmetry, i.e. in the zero limit of the trapping. When the symmetry is broken, the snake instability starts to develop at the location(s) of the symmetry breaking. Note that the symmetry breaking is not spontaneous, but induced.

\begin{figure}
\centering
\includegraphics[width=0.45\textwidth] {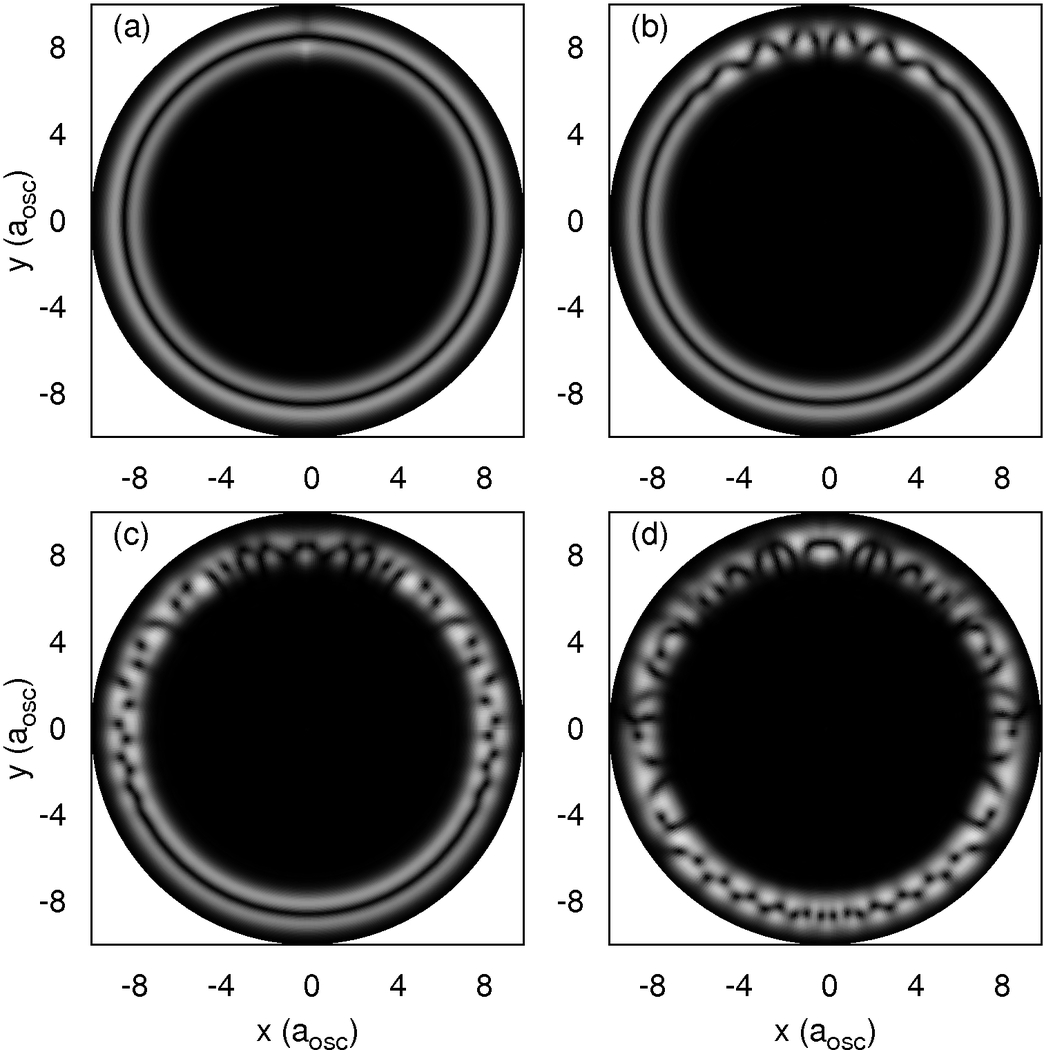}   
\caption{\label{fig:snakeinstabpolar} Snapshots of real-time propagation of the GPE~\eqref{GPE2Dapolar} in a polar grid at (a) $t = 0.0$, (b) $t = 1.5$, (c) $t = 3.0$, and (d) $t = 4.4/ \omega_x$ with the harmonic torus~\eqref{SonT:1} with $k = 96$, $R_S = 8.5$, and $C_{2D} = 400$, but with a slight added anisotropy at $\theta = \pi/2$. The transverse oscillations of the snake instability are seen to start immediately originating at the trap anisotropy and propagate along the ring with a finite speed, in accordance with the model discussed in the text.}
\end{figure}

\begin{figure}
\centering
\includegraphics[width=0.45\textwidth] {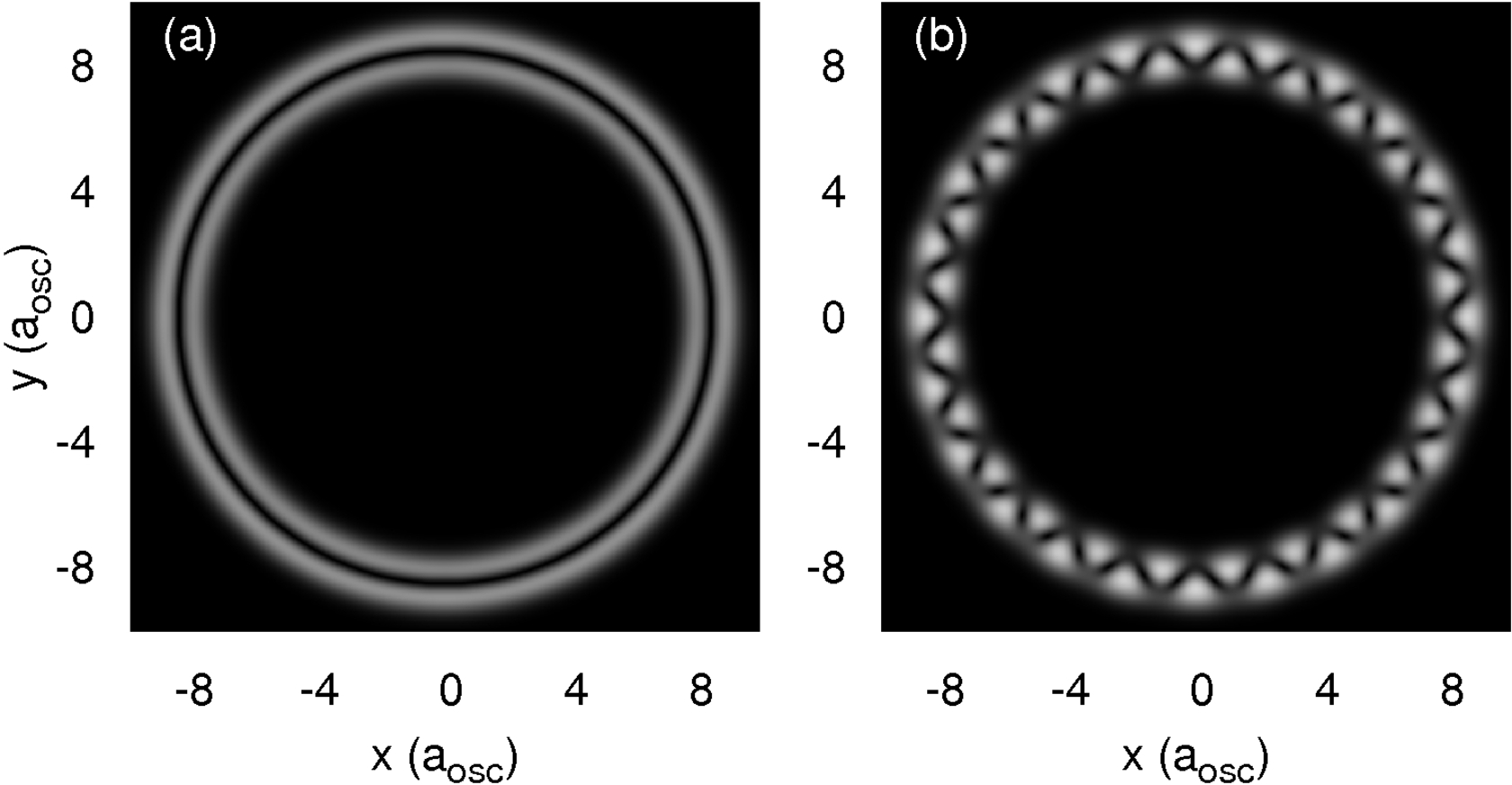}   
\caption{\label{fig:Snake_Instab_Torus} Snapshots of real-time propagation of the GPE~\eqref{GPE2Da} in a rectangular grid at (a) $t = 0.0$ and (b) $t = 2.8/ \omega_x$ with the same parameters of Fig.~\ref{fig:snakeinstabpolar} excluding the anisotropy. The transverse oscillations of the snake instability are seen to originate over the whole ring resulting in the collapse of the ring dark soliton (producing eventually a necklace of vortex-anti-vortex pairs).}
\end{figure}

On the torus geometry, we have in general observed the snake instability provided we use polar coordinates and an asymmetric potential (see Fig.~\ref{fig:snakeinstabpolar}) or rectangular coordinates (see Fig.~\ref{fig:Snake_Instab_Torus}), but the trap parameter $k$, the nonlinear coupling constant $C_{2D}$, the soliton radius $R_S$, and the particular way in which the symmetry is broken affect the decay time. To eliminate motional dynamics we take into account the effective curvature-induced potential, and provided the radius of the ring is large enough, the stationary radius is determined by~\cite{PhysRevLett.90.120403}
\begin{equation}
\label{eqn:statrad}
-\frac{1}{2}(V_{\mathrm{trap}})_R + \frac{1}{3R} = 0.
\end{equation}
Our simulations show that the higher the curvature of the trap (in the radial direction), the faster is the onset of the snake instability. Also, the decay time, $T_d$, decreases if we increase the nonlinearity while keeping the other parameters fixed. In a fixed harmonic trap~\eqref{SonT:1} with $k = 1$ and $R_S = 14.5$, we find that $T_d \sim C_{2D}^{-0.55}$. 

For a fixed nonlinearity of $C_{2D} = 400$, the ring dark soliton survives for $\approx 10$ times longer in a harmonic trap ($R_S = 0$) than in a torus of similar size. The survival times for the $R_S = 0$ case are much longer because the ring dark soliton can move quite freely, whereas toroidal confinement restricts motion, which leads to faster decay. If the soliton moves fast enough, it will fade out without the snaking instability. The effect of the motion is in line with the results of the experiments for moving planar dark solitons~\cite{PhysRevLett.83.5198, ZacharyDutton07272001}. Furthermore, if the soliton is stationary, also the vortex-anti-vortex necklace will be stationary. 

We have also observed periodic revival and subsequent decay into a vortex-anti-vortex necklace of the ring dark soliton in the harmonic torus~\eqref{SonT:1} with $k = 1$, $R_S = 7.5$, and $C_{2D} = 400$. The initial radius of the soliton is given by Eq.~\eqref{eqn:statrad}. The initial ring dark soliton decays to a vortex-anti-vortex necklace, but is revived several times later on with a period of $\approx 8.8 / \omega_x$.

One should note that tightening the trap or increasing the nonlinearity have a similar effect on the healing length,
\begin{equation}
\label{HealL}
\xi^2 = \frac{1}{8\pi n_0 a},
\end{equation}
where $n_0$ is the bulk density. It characterises both the width of the density notch in a dark soliton as well as the size of vortex cores. Also, $\xi$ will decrease if we make the trap tighter (to preserve normalisation) or increase the nonlinearity. With decreasing healing length it becomes easier for the snake instability to deform the condensate with respect to the bulk.

\begin{figure}
  \centering
  \includegraphics[width=0.45\textwidth]{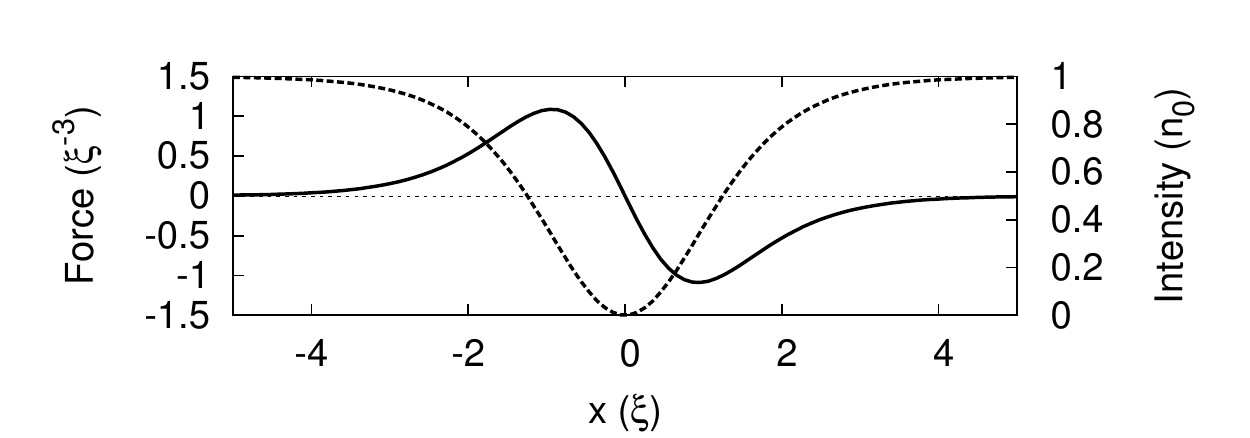}
  \caption{The hydrodynamical force (solid), Eq.~\eqref{SonT:4b}, around the notch of a planar dark soliton at $x = 0$ (dashed). Notice the sign of the force and that the force scales as $\sim \xi^{-3}$, explaining the relationship between the decay time and $\xi$.}
  \label{fig:hyddyn}
\end{figure}

Of course, nothing would happen if there was no driving force trying to deform the condensate around the soliton's notch in the first place. As a first approximation, this force can be understood if we replace a homogeneous bulk condensate with an infinite potential for $x < x_S$, where $x_S$ marks the position of the wall. The infinite wall forces a zero boundary condition for the wavefunction just like a planar dark soliton at $x_S$, so for $x > x_S$ nothing changes. This happens because the kink
\begin{equation}
\label{SonT:4}
\psi = \tanh{\left( \frac{x}{\sqrt{2}}\right) } \qquad (x > 0)
\end{equation}
solves Eq.~\eqref{GPE2Da} ($C_{2D} = 1$) with an infinite wall potential at $x < 0$ and a zero potential elsewhere (uniform for $y$ and $z$). Therefore, the dark soliton strip can be thought of as serving the role of a two-sided wall, one side facing towards increasing coordinate and the other side in the opposite direction. Sudden removal of such a wall potential will result in the expansion of the condensate to the previously forbidden area. This can be seen from the hydrodynamical description of the condensate ($\psi = \sqrt{n}e^{iS}$, where $S$ is the phase):
\begin{equation}
\label{SonT:4a}
m \frac{d\mathbf{v}}{dt} = -\frac{1}{n}\nabla p - \frac{m}{2} \nabla v^2 + \nabla \left( \frac{\hbar^2}{2m \sqrt{n}}\nabla^2 \sqrt{n} \right) - \nabla V_{\mathrm{trap}},
\end{equation}
where $p$ is the pressure, $\textbf{v} = \frac{\hbar}{m} \nabla S$ the fluid velocity, and $n$ is the (local) density. Since the length scale of spatial variation of the condensate wavefunction at the soliton's notch is of the order of $\xi$, the relevant and dominant term is the quantum pressure. Focusing on the planar case, and measuring length, time, and density in terms of $\xi$, $\frac{2m}{\hbar}\xi^2$, and $n_0/N$, respectively ($\xi = \hbar = 2m = n_0/N = 1$) the quantum pressure term in Eq.~\eqref{SonT:4a} reduces to
\begin{equation}
\label{SonT:4b}
\frac{1}{2}\frac{d\mathbf{v}}{dt} =  -\nabla \left( \frac{1}{\sqrt{n}}\nabla^2 \sqrt{n} \right),
\end{equation}
where the extra minus sign follows from the soliton wavefunction ($S \to -S$). Equation~\eqref{SonT:4b} is plotted for a dark soliton strip $n = \tanh^2{\left( x/\sqrt{2}\right) }$ in Fig.~\ref{fig:hyddyn}.
The force is pointing towards the notch on both sides, and in particular, it scales as $\sim \xi^{-3}$. Filling the notch with a repulsive component~\cite{0953-4075-44-19-191003} reduces the effect of this force, explaining why it acts as a stabilising factor.

Provided this force can point at least infinitesimally askew, the two sides will then eventually slip past each other, twisting the density notch of the soliton as they expand. Since there is a phase difference of $\pi$ between the two sides, twisting the density notch by such an expansion will result in an accumulation of phase by $2\pi$ when we go around one dot-like density notch the twisting leaves, i.e. vortices. The phase singularity generates a topological protection for the perseverance of the vortices. This process will then propagate along the ring (see Figs.~\ref{fig:snakeinstabpolar} and~\ref{fig:Snake_Instab_Torus}).

\begin{figure}
  \centering
  \includegraphics[width=0.45\textwidth]{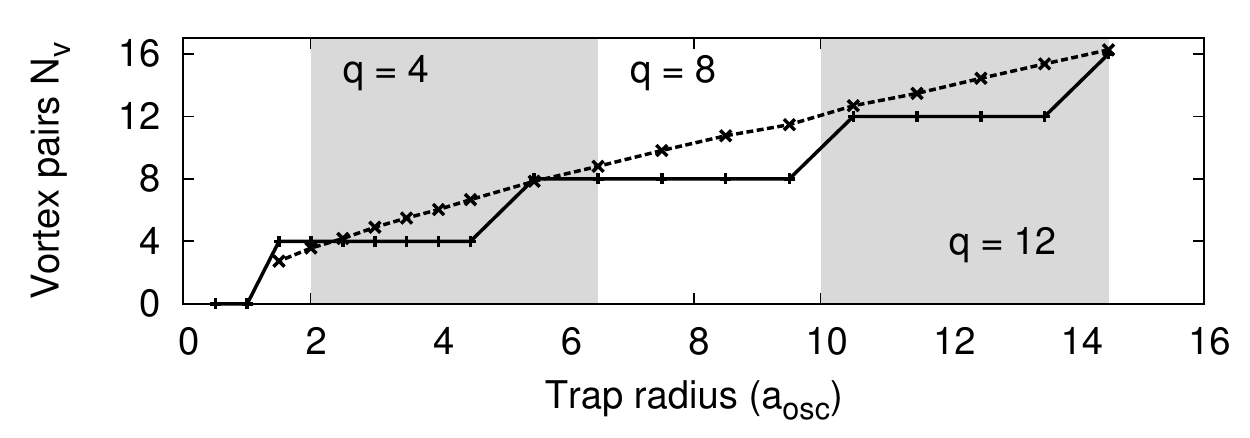}
  \caption{The correspondence between the predicted (dashed, see Eq.~\eqref{SonT:5}) and measured (solid) number of vortex-anti-vortex pairs for the harmonic potential~\eqref{SonT:1} with $k = 1$ and $C_{\mathrm{2D}} = 400$. The healing length was measured by finding a best fit to $\frac{\mu-\frac{1}{4}(R-R_S)^2}{C_{\mathrm{2D}}} \tanh^2{\left(\frac{R-R_S}{\sqrt{2} \xi}\right)}$. For the low radii the ring soliton disappears as sound waves without vortices. The shaded areas show where the complex Bogoliubov modes such that $q$ is a multiple of 4 are active. In the middle of the regions the modes are also the fastest ones (see Fig.~\ref{fig:BdGE_torusharm}), but they are still chosen by the condensate near the region boundaries where other modes would be faster as such, because $N_v$ must be a multiple of four by symmetry reasons.}
  \label{fig:N_v}
\end{figure}

The number of vortex-anti-vortex pairs is then determined by the length of the soliton and the size of the vortex core. As a first approximation we can say that the number of vortex-anti-vortex pairs, $N_v$, will be
\begin{equation}
\label{SonT:5}
N_v \sim \frac{2\pi R_S}{8\xi} = \sqrt{\frac{\pi^3 a n_0}{2}} R_S 
\end{equation}
naturally such that $N_v$ is an integer (see Fig.~\ref{fig:N_v}) because each vortex-anti-vortex pair and the distance to the next pair needs a total length of $\sim 8\xi$ on the circumference. Their locations within the ring is determined by where the symmetry was broken.

\begin{figure}
  \centering
  \includegraphics[width=0.45\textwidth]{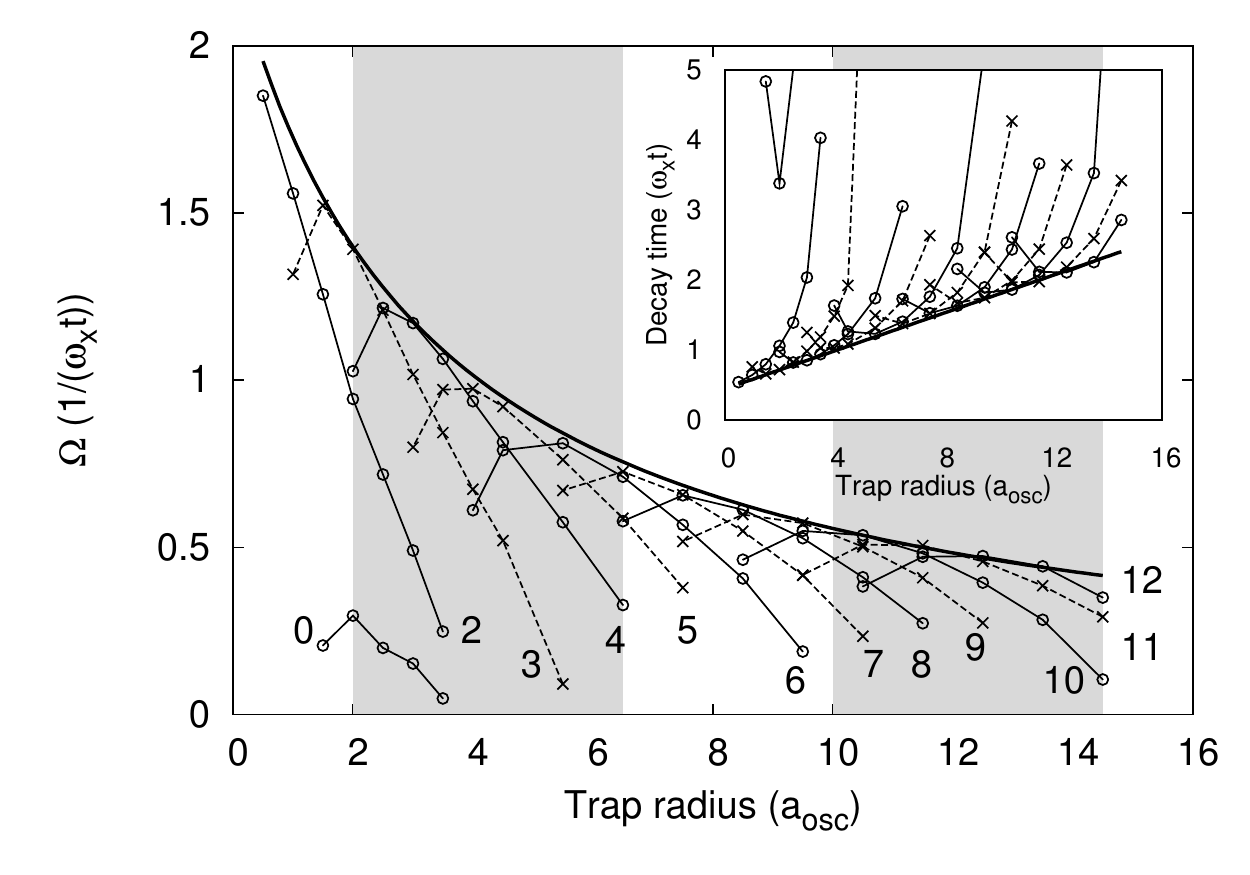}
  \caption{Results of the Bogoliubov analysis for the harmonic potential~\eqref{SonT:1} with $k = 1$ and $C_{\mathrm{2D}} = 400$ such that the soliton is printed at the radius given by Eq.~\eqref{eqn:statrad}. Shown are the Bogoliubov eigenvalues $\Omega$ for modes with a dynamical instability, labelled by the value of $q$. There is only at most one imaginary eigenvalue per $q$. The inset shows the reciprocal numbers giving the decay times. The solid black line without markers is given by $\Omega = 1.0/(0.135R+0.444)$. For low radii the primary decay channels of the RDS are the $q = 2, 3$ modes, while higher modes are dynamically stable. As the radius is increased the higher modes become unstable as well. The shading is the same as in Fig.~\ref{fig:N_v}. When there is no soliton all the eigenvalues are real.}
  \label{fig:BdGE_torusharm}
\end{figure}

The Bogoliubov decay times define a quantitative time scale related to the snake instability (see Fig.~\ref{fig:BdGE_torusharm}). According to these times, a larger radius leads to a longer decay time, which can be understood because in a larger trap the bulk density is reduced to maintain normalisation. This increases $\xi$ and hence the decay time.

\begin{figure}[ht]
  \centering
  \includegraphics[width=0.45\textwidth]{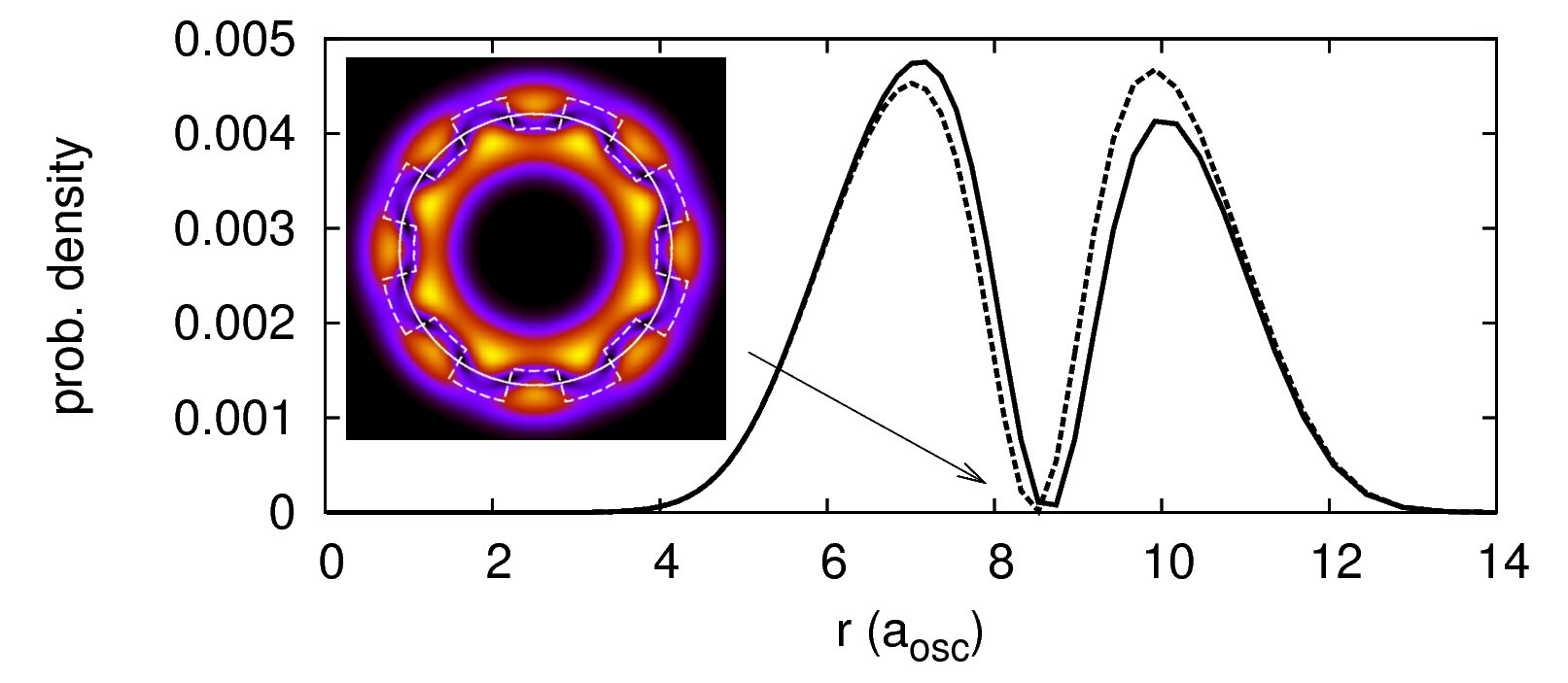}
  \caption{(Colour online.) The wave functions $|\phi_0|^2$ (solid) and $|\phi_0 + u + \bar{v}|^2$ (dashed) at $\theta = 2\pi/8$ for a ring dark soliton in the harmonic trap~\eqref{SonT:1} with $k = 1$ and $C_{\mathrm{2D}} = 400$ such that the soliton is printed at $R = 8.65407$ (as given by Eq.~\eqref{eqn:statrad}). $u$ and $v$ correspond to the only imaginary Bogoliubov eigenvalue $\Omega = 0.613i$ of the mode $q = 8$. The inset shows how the location of the soliton's notch is affected by the Bogoliubov mode around the circumference, mimicing the snake instability and coinciding exactly with the locations of the vortices (the Bogoliubov mode in the inset is exaggerated for clarity). For this radius, the $q = 8$ mode is the fastest, and it produces $q = N_v = 8$ pairs (see Fig.~\ref{fig:N_v}), in accordance with observations.
}
  \label{fig:BdGE_snakemodes}
\end{figure}

Because we have observed the snaking instability in general, and because there is only one imaginary Bogoliubov mode (per $q$), it is reasonable to assume they are related. The connection between the form of the complex Bogoliubov modes and the planar snake instability has been discussed in Ref.~\cite{PhysRevA.62.053606}, and here we report similar results for the ring and elucidate further the role of the complex modes. Firstly, we note that for an exact soliton-like dark ring solution of Eq.~\eqref{GPE2Da}, the complex Bogoliubov modes are very similar to Fig.~\ref{fig:BdGE_torusharm}, and show transverse oscillations around the ring resembling the snake instability~\cite{ths2012}. This is the case for the numerically imprinted ring dark soliton as well (see Fig.~\ref{fig:BdGE_snakemodes}). 

In particular, the number of vortex-anti-vortex pairs, $N_v$, is determined by the decay channels, the channel $q$ giving $q$ pairs (see Fig.~\ref{fig:N_v}). Due to the symmetries of the rectangular square grid, $N_v$ must be a multiple of four, but also the fastest decay channels (i.e. largest imaginary eigenmodes) will be preferred. Furthermore, Eq.~\eqref{SonT:5} sets an (approximate) upper bound, and of course the way in which the symmetry is broken dictates how the snake instability can take place. The exact choice of $q$s is then a balance of several factors.

In summary, we have studied the stability of ring dark solitons on the torus geometry, elucidating the connection between relevant symmetries, the BdG spectrum and the snake instability. In addition to assessing the role of toroidal confinement in ring soliton decay times, we have provided a physical description for the decay of dark solitons into vortices. Even though we have considered a ring dark soliton, it is important to note that our modelling is general in nature. The location(s) of the symmetry breaking are seen to be the origins of the snake instability in many other configurations as well~\cite{ZacharyDutton07272001,Kamchatnov20112577,PhysRevLett.97.180405}.

\begin{acknowledgments}
We acknowledge the support of the Academy of Finland (grant 133682), the Finnish Cultural Foundation (LT), and Jenny and Antti Wihuri Foundation (LT). We also thank Jarmo Hietarinta for fruitful discussions.
\end{acknowledgments}

\bibliographystyle{apsrev4-1}
\bibliography{references}
\end{document}